\documentclass[11pt]{article}
\usepackage[margin=0.88in]{geometry}
\usepackage{amsmath,amssymb}
\usepackage{booktabs}
\usepackage{array}
\usepackage{enumitem}
\usepackage{microtype}
\usepackage{tabularx}
\usepackage{tikz}
\usetikzlibrary{arrows.meta,positioning,shapes.geometric,fit,calc,plotmarks}
\usepackage{pgfplots}
\pgfplotsset{compat=1.18}
\usepackage[hidelinks]{hyperref}
\usepackage{url}

\hypersetup{
  pdftitle={Assurance Envelopes for Autonomous Coding Agents: Minimum-Cost Evidence for Software Change},
  pdfauthor={Anjan Goswami},
  pdfsubject={Minimum compositional support for supplied software-change obligations},
  pdfkeywords={autonomous coding agents, software assurance, agent context selection, evidence selection, software evolution}
}

\newcommand{\method}[1]{\textsc{#1}}
\newcommand{\allmethod}{\method{all}}
\newcommand{\flatmethod}{\method{flat}}

\title{\textbf{Assurance Envelopes for Autonomous Coding Agents:\\Minimum-Cost Evidence for Software Change}}
\author{Anjan Goswami\\SmartInfer AI}
\date{September 2026}

\begin{document}
\maketitle

\begin{abstract}
When a coding agent returns to existing software, it inherits evidence from earlier
engineering work: passing tests, type checks, discharged contracts, static analyses, and
monitored traces. Reloading all of it is wasteful, but dropping a piece the change
depends on can leave a required property unsupported. Given the properties a change must
preserve, its \emph{obligations}, we ask which least-cost subset of the available
evidence is enough to re-establish them, and we call such a subset a
\emph{task-conditioned assurance envelope}. We make the composition explicit: evidence
and the rules that combine it form a typed inference graph, an obligation is met when
forward chaining from the selected evidence reaches it, and we validate every selection
by that closure rather than by trusting the optimizer.

We evaluate the formulation in two stages. The software-derived graphs in our evaluation
come from preserved outcomes of prior AI coding-agent runs; this study freezes those
artifacts and asks which accumulated assurance evidence should be restored for a
subsequent task. Small graphs hand-specified from preserved Rust, IronBlocks, and Pong
outcomes show that the minimum envelope depends on the task,
that no sufficient envelope may exist when current evidence cannot re-establish a required property,
that some properties need several pieces of evidence together, and that expanding the
requirements adds evidence rather than replacing it. A prespecified synthetic benchmark
of 249 graph instances then characterizes computation. A baseline that discards the
``several pieces together'' structure necessarily fails to re-derive those properties, a
consequence we quantify; every exact cross-check that completed agreed with the CP-SAT
optimizer; and median solve time stayed below 20 ms at 500-evidence graphs, except that
graphs with many alternative derivations per target timed out at far smaller sizes,
showing that derivation structure rather than raw size drives difficulty. The
contribution is a bounded application of established optimization to selecting assurance
context for a supplied software-change task; automatic discovery of the obligations and
downstream agent benefit remain open.
\end{abstract}


\section{Introduction}

A coding agent may return to a repository that earlier runs have built, tested,
analyzed, and modified. The engineering knowledge it inherits is distributed
across type checks, contracts, tests, static analyses,
trace monitors, operational policies, and requirement history. Source code
alone does not reconstruct those conclusions. Conversely, supplying every old
artifact and tool result for every change is neither necessary nor obviously
useful.

This setting is concrete in our evaluation: the software artifacts underlying the
preserved Rust, IronBlocks, and Pong outcomes were produced in prior AI coding-agent
runs, primarily Claude- and Codex-based. We study which assurance evidence from those
runs should be restored for a subsequent task, rather than how the original code was
generated.

The relevant context depends on the requested change. A local refactoring may
need build and interface obligations (the properties it must preserve), while a change
to cancellation behavior may also need a trace obligation about events that can occur
later. Selecting
too much evidence carries irrelevant history. Selecting too little is more
serious: an omitted premise can make a claimed obligation unsupported.
Independent test coverage does not capture this selection problem: some
conclusions require several premises together, while others have alternative
derivations. For a long-horizon coding agent, the problem is therefore not only what
repository text to retrieve, but which prior machine-produced engineering evidence must
be restored so the supplied task obligations remain supported.

This paper studies the problem after the task obligations have been supplied.
We call such a sufficient selection, chosen for one program, task, and context, a
\emph{task-conditioned assurance envelope}: the least-cost bundle of existing evidence
that is enough to re-establish the properties the change must preserve. The primary research
question is:

\begin{quote}
\textbf{RQ.} Given a program artifact, a software-change task, and a
semantic/verification context, which minimum-cost subset of the available typed
evidence derives every supplied task obligation under the explicit composition
rules?
\end{quote}

We take the task obligations and applicable inference rules as inputs rather
than infer them from natural language or code; the problem studied here is to
select and validate their evidence support.

Independent evidence coverage reduces to weighted set cover, while conjunctive
premises, alternative derivations, and shared support lead to known
minimum-support problems over directed hypergraphs or AND/OR graphs
\cite{feige,gallo,eiter}. Assurance cases and lifecycle systems already
structure and maintain evidence \cite{gsn,sacm,opencert,amass,access}. Against
this background, we study one operational specialization: selecting a minimum
support for the supplied obligations of a future coding task.

We make four contributions:

\begin{enumerate}[leftmargin=1.5em]
\item \textbf{Formulation.} We formulate the evidence needed for one
software-change task as sufficient support over typed, artifact- and
context-indexed evidence and explicit inference rules. We relate this
formulation to known cover and proof-support problems.
\item \textbf{Provenance-derived evaluation.} We evaluate assurance graphs
hand-specified from preserved Rust and IronBlocks outcomes, demonstrating
task-dependent minimum support, infeasibility when current evidence cannot
derive a required root, and a conjunctive dependency erased by a prespecified
pairwise baseline.
\item \textbf{Requirement evolution.} We evaluate requirement expansion in a
Pong graph hand-specified from preserved outcomes: the prior support remains
applicable, but one new evidence leaf is needed for the expanded requirement
set.
\item \textbf{Implementation and scalability.} We implement deterministic
graph validation and closure, an exhaustive oracle, and CP-SAT optimization. We
use a prespecified 249-instance benchmark to cross-check the optimizer and
characterize bounded computational behavior.
\end{enumerate}

Together, these contributions provide a formalization and bounded evaluation
over small software-derived graphs and a synthetic scale study; task roots and
inference rules are supplied.

\section{Task-Conditioned Assurance Envelopes}

Before introducing notation, we fix what the pieces are in ordinary engineering terms.
Each formal object below answers a concrete question about a software change.

\paragraph{Obligations: the properties a change must preserve.}
When a change lands, some properties still have to hold: the build succeeds, an
interface contract is met, a test passes, a concurrency invariant is maintained, or a
rejected action never commits. We call each such property an \emph{obligation}.
Obligations are heterogeneous: a compiler establishes a type judgment, a test
establishes behavior on selected inputs, a contract proof establishes a local state
relation, a monitor establishes a property over bounded execution traces, and an
operational rule concerns whether an action crosses a commitment boundary. These are
different kinds of guarantee, not points on one strength scale, so evidence for one
obligation does not substitute for another unless an explicit rule connects them.

\paragraph{Context: the conditions under which evidence holds.}
What a piece of evidence actually establishes depends on conditions we bundle as the
\emph{context}: the compiler and its options, the language profile, the runtime and
memory model, the dependencies, and the analyses actually run. A language name alone
does not fix them. In a pinned safe-Rust profile, ownership and typing discharge some
obligations while unsafe code, foreign-function calls, and trace properties remain
external \cite{rustbelt}; C/C++ concurrency requires an explicit memory-model treatment
\cite{batty}; concurrent code needs temporal properties that local
pre/postconditions do not give \cite{pnueli,linearizability}. The same evidence can be
decisive in one context and irrelevant in another.

\paragraph{Evidence, rules, closure, and sufficiency.}
\emph{Evidence} is what a prior tool run left behind: a passing test, a type-check
result, a discharged proof obligation, a static-analysis verdict, a monitored trace
outcome. Some obligations follow from a single piece of evidence; others hold only when
several pieces hold together, and some can be reached by more than one route. We record
these dependencies as an \emph{inference graph}: a directed graph in which one step may
require several inputs at once (an AND, such as ``detection and mediation together imply
commit-safety'') and in which one obligation may have alternative supporting steps (an
OR). Given a chosen set of evidence, its \emph{closure} is everything the rules let you
conclude by applying them repeatedly until nothing new appears. The chosen evidence is
\emph{sufficient} for a task when its closure contains every obligation the task
requires; when no available evidence is sufficient, no assurance envelope exists for the supplied obligations under the current evidence and rules. We
take the task's obligations and the applicable rules as given and study which evidence
to select; discovering the obligations is a separate problem that we do not solve here.

\subsection{Problem objects}

We now make these objects precise. Let $P$ be an exact artifact identity, $t$ a
requested engineering task, and $M$ the context above. Let
\[
  O(P,M)
\]
be the universe of obligations that could be required, and let
\[
  R(P,t,M) \subseteq O(P,M)
\]
be the obligations this task actually requires, its \emph{roots}. This subset is the
conditioning decision: a change does not necessarily require every known property.
Constructing a complete $R(P,t,M)$ remains an external requirements and change-impact
problem.

Let $E$ be the available evidence and $G$ the finite typed inference graph. An evidence
item directly establishes selected obligations. A graph edge maps a set of antecedent
obligations that must all hold to one consequent obligation, and several edges sharing a
consequent are alternative derivations. (Formally $G$ is a directed hypergraph; we call
it the inference graph throughout.)

\begin{figure}[t]
\centering
\begin{tikzpicture}[
  node distance=10mm and 11mm,
  box/.style={draw,rounded corners,align=center,minimum height=9mm,minimum width=27mm,fill=blue!5},
  arrow/.style={-{Latex[length=2mm]},thick}
]
\node[box] (task) {artifact $P$, task $t$\\context $M$};
\node[box,right=of task] (roots) {supplied roots\\$R(P,t,M)$};
\node[box,right=of roots] (graph) {typed inference\\graph $G$};
\node[box,right=of graph] (env) {minimum sufficient\\support $S^*$};
\draw[arrow] (task) -- (roots);
\draw[arrow] (roots) -- (graph);
\draw[arrow] (graph) -- (env);
\end{tikzpicture}
\caption{A task-conditioned envelope selects evidence and derivations for
supplied roots. Root discovery is outside the solved problem.}
\label{fig:flow}
\end{figure}
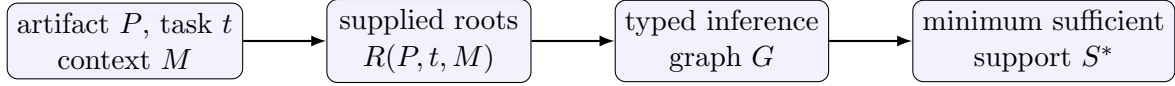

\subsection{Derivation and sufficiency}

Formalizing the closure described above, for $S\subseteq E$ we define $Cn_G(S)$ by
deterministic forward chaining. It
initially contains every obligation directly established by selected evidence.
It then repeatedly adds a rule's consequent when all antecedents are supported,
until no new obligation is added. In V1, the graph is finite, monotone, and
acyclic.

A selection is sufficient exactly when
\begin{equation}
  R(P,t,M) \subseteq Cn_G(S).
  \label{eq:sufficient}
\end{equation}
Operationally, Eq.~\ref{eq:sufficient} is the acceptance criterion: every
required root must have a derivation from selected evidence. The independent
closure check validates solver-reported feasibility.

\begin{figure}[t]
\centering
\begin{tikzpicture}[
  evidence/.style={draw,rounded corners,fill=green!12,minimum width=13mm,minimum height=7mm},
  obligation/.style={draw,circle,fill=blue!8,minimum size=8mm,inner sep=1pt},
  rule/.style={draw,diamond,fill=orange!15,aspect=1.7,inner sep=1.5pt},
  arrow/.style={-{Latex[length=1.8mm]},thick},
  node distance=8mm and 12mm
]
\node[evidence] (ea) {$e_a$};
\node[evidence,below=of ea] (eb) {$e_b$};
\node[obligation,right=of ea] (a) {$a$};
\node[obligation,right=of eb] (b) {$b$};
\node[rule,right=18mm of $(a)!0.5!(b)$] (r) {$r_\wedge$};
\node[obligation,right=of r] (c) {$c$};
\node[evidence,above=of c] (ec) {$e_c$};
\draw[arrow] (ea)--(a);
\draw[arrow] (eb)--(b);
\draw[arrow] (a)--(r);
\draw[arrow] (b)--(r);
\draw[arrow] (r)--(c);
\draw[arrow] (ec)--(c);
\node[draw,dashed,rounded corners,fit=(ea)(eb)(a)(b)(r),inner sep=3mm,label=below:{AND support}] {};
\end{tikzpicture}
\caption{A worked inference graph. The rule node preserves $a\land b\Rightarrow
c$; direct evidence $e_c$ supplies an alternative route. Costs determine which
valid support is selected.}
\label{fig:worked}
\end{figure}
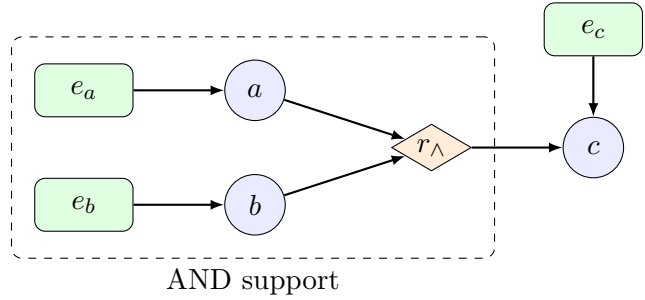

Consider evidence $e_a$ and $e_b$ that establish $a$ and $b$, with a rule
$a\land b\Rightarrow c$. For task root $c$, neither singleton is sufficient;
$\{e_a,e_b\}$ is. If a separate evidence leaf $e_c$ directly establishes $c$,
then the graph has an OR alternative. Figure~\ref{fig:worked} shows this small
support problem.

\subsection{Minimum support, ties, and infeasibility}

For scalar additive leaf cost, a minimum-cost envelope is
\begin{equation}
  S^*(P,t,M) \in
  \arg\min_{S\subseteq E} Cost(S)
  \quad\text{subject to}\quad
  R(P,t,M)\subseteq Cn_G(S).
  \label{eq:opt}
\end{equation}
Equation~\ref{eq:opt} asks for the least-cost evidence set that derives every
required root under the declared derivation semantics. Multiple sets may attain
the same minimum. If none derives every root, current evidence is insufficient
for the task.

Inclusion minimality and minimum cost are distinct. The experiments use unit
cost as the primary measure of support cardinality and a small deterministic
integer-cost secondary condition. Multiobjective choices involving latency,
freshness, human work, or risk require a policy and possibly a Pareto frontier;
they are not part of V1.

Under independent fixed coverage, Eq.~\ref{eq:opt} is weighted set cover. With
conjunctive rules, intermediate obligations, alternatives, and shared leaves,
it is a minimum proof/support problem over a directed hypergraph. Both are
known optimization settings \cite{feige,gallo,eiter}; the task-conditioned
software-assurance use is the subject of this paper.

\section{Provenance-Derived Software Graphs}

This section evaluates assurance graphs hand-specified from preserved outcomes of prior
AI coding-agent runs. We freeze the generated software and its recorded engineering
evidence; the present experiment studies task-conditioned evidence support rather than
builder quality. It does not rerun Rust, IronBlocks, or Pong. Obligations,
rules, and unit costs were prespecified before graph evaluation. Exact
exhaustive search produced the results in Table~\ref{tab:software}, and the
independent closure validator checked each selected support.

\begin{table}[t]
\centering
\small
\setlength{\tabcolsep}{3.5pt}
\begin{tabularx}{\linewidth}{p{0.18\linewidth} p{0.25\linewidth} c X}
\toprule
\textbf{Graph/task} & \textbf{Required support} & \textbf{Cost} & \textbf{Result} \\
\midrule
Rust C, local & $e_{build,C},e_{local,C}$ & 2 & feasible; trace evidence omitted \\
Rust C, trace & $e_{build,C},e_{local,C},e_{trace,C}$ & 3 & feasible; task adds trace root \\
Rust B, local & $e_{build,B},e_{local,B}$ & 2 & feasible \\
Rust B, trace & none & -- & no sufficient envelope; no positive trace support \\
IronBlocks & $e_{detect},e_{mediated}$ & 2 & unique compositional optimum; FLAT chooses invalid cost-1 singleton \\
Pong, $R_0$ & $e_{prior,post}$ & 1 & prior repaired-artifact obligation \\
Pong, $R_1$ & $e_{prior,post},e_{horizontal,post}$ & 2 & retains one leaf and adds one \\
\bottomrule
\end{tabularx}
\caption{Exact results for the prespecified graphs hand-specified from preserved
software outcomes. Dashes denote an infeasible support problem (no sufficient envelope), not a solver error.}
\label{tab:software}
\end{table}

\subsection{Rust: task conditioning and infeasibility}

The preserved Rust calibration evidence distinguishes build, local-state, and
bounded-trace obligations. Local evidence covers all 30 prespecified
state/event pairs. Trace evidence covers 781 event sequences through depth
four. There is no inference rule from local success to trace success.

For Fixture C, a local task requires build and local evidence, giving the
unique cost-2 support $\{e_{build,C},e_{local,C}\}$. Adding the trace root
changes the exact optimum to cost 3 and adds $e_{trace,C}$. The artifact and
available graph are the same; only the supplied task requirements differ.

Fixture B has positive build and local evidence but failed trace evidence. Its
local task is feasible at cost 2. Its trace task has no sufficient envelope because a failed
check is retained as provenance but does not establish the positive trace
obligation. Task conditioning can therefore change the minimum envelope or
reveal that none exists. These results are computed over preserved,
depth-bounded fixture outcomes; they neither rerun Rust nor establish an
unbounded temporal property.

\subsection{IronBlocks: compositional support}

The IronBlocks graph contains two operational premises:
\[
  \phi_{detect}\land\phi_{mediated}
  \Longrightarrow \phi_{commit\_safe}.
\]
Detection evidence says that an invalid proposal is recognized under the
prespecified action semantics. Mediation evidence says that this decision governs
submission at the recovered commitment boundary. Under the supplied rule, the
unique compositional optimum requires both leaves and costs 2.

The \flatmethod{} baseline mechanically replaces an AND rule with independent
pairwise links. It consequently reports two cost-1 optima, one for each leaf.
The independent closure check rejects both because neither derives commitment
safety alone. The composition relation was supplied by construction, so the
graph evaluation measures the consequence of preserving or erasing it.

The preserved historical experiment used matched synthetic proposal streams
and 360 runs per configuration. Its gate behavior is partly by construction,
as a correctly mediated rejection should not cross the commitment boundary.
Runtime enforcement and the distinction between observation and intervention
are established prior art \cite{schneider,editautomata,falcone}. Here they
provide a concrete conjunction in an assurance-support graph.

\subsection{Pong: requirement evolution}

Pong supplies a different phenomenon: the selected requirement set expands.
The graph fixes the repaired artifact and compares
\[
  R_0=\{\phi_{prior}\}
  \quad\text{with}\quad
  R_1=R_0\cup\{\phi_{horizontal}\}.
\]
The exact $R_0$ envelope has cost 1 and contains $e_{pong\_prior\_post}$. The
$R_1$ envelope retains that leaf and adds $e_{pong\_horizontal\_post}$, giving
cost 2. The old evidence remains applicable, but the old envelope is
insufficient for the expanded requirement set.

The horizontal requirement is
\[
  \frac{|v_x|}{\sqrt{v_x^2+v_y^2}} \geq 0.5.
\]
Retained documentation reports 0.481 before repair and 0.500 after repair,
while the previous 21-test suite remained green. The pre- and post-repair source
revisions and committed oracle are preserved. The numerical values are retained
through documentation and sanitized/reconstructed evidence rather than a fresh
whole-artifact reproduction. The graph fixes the repaired artifact and
evaluates requirement expansion, not artifact-change invalidation.

\section{Scalability Benchmark}

\subsection{Prespecified design}

The scalability protocol was prespecified and frozen before generation and execution. It
defines 249 graph/cost instances, 267 graph/task evaluations, and 801 method
rows. Three fixed seeds instantiate eight deterministic families:

\begin{itemize}[leftmargin=1.5em]
\item F1 independent cover;
\item F2 conjunctive chains;
\item F3 alternative derivations;
\item F4 shared support;
\item F5 irrelevant-evidence noise;
\item F6 mixed composition;
\item F7 constructed infeasibility; and
\item F8 nested requirement expansion.
\end{itemize}

Primary evidence sizes are 10, 25, 50, 100, 250, and 500. Exact-only points
add sizes 15 and 20. Every feasible graph is built from a known support
skeleton before alternatives or distractors are added. F7 removes prespecified
establishment edges from an F2 skeleton. Primary evidence cost is one; six
secondary instances use deterministic integer costs from one through five.

The three methods are:

\begin{description}[leftmargin=2.3cm,style=nextline]
\item[ALL] retain every evidence leaf and evaluate closure;
\item[FLAT] project every AND rule into independent pairwise
antecedent-to-consequent links, then solve weighted set cover; and
\item[COMPOSITIONAL] solve minimum-cost support on the original graph.
\end{description}

\flatmethod{} is intentionally lossy but mechanically defined before results.
It represents a plausible structure that preserves pairwise ``supports'' links
while omitting that premises must be combined. It is a diagnostic baseline,
not a claim about every non-hypergraph assurance representation.

\subsection{Optimization and validation}

NetworkX stores the inference graph and computes deterministic closure. OR-Tools
CP-SAT, a constraint-programming SAT-based optimizer, uses binary variables for
evidence, obligations, and rules, with
constraints that encode conjunction and require every supported obligation to
have a justification. Runs use one worker, seed zero, presolve, no hints, and a
60-second per-method timeout.

For every graph with at most 20 evidence leaves, independent exhaustive solvers
enumerate all subsets for both the compositional and flat semantics. Before the
full run, the mandatory gate required all 21 prespecified n=10 F1--F7 instances
to agree with CP-SAT on feasibility, objective, and valid support. The
independent closure validator separately checked every CP-SAT support. Timeouts
and infeasibility remained in the data.

\section{Scalability Results}

\subsection{Loss from erasing conjunction}

Because \flatmethod{} discards the requirement that several premises hold together,
it cannot derive any obligation that depends on a conjunction; the benchmark quantifies
how often that matters. \flatmethod{} reported coverage that closure under the original
graph rejected in 135 of 267 task evaluations (3,765 required roots). Every such result
fell in a family containing required conjunctions (F2, F4, F6, F7), and there was none
in F1, F3, F5, or F8. Table~\ref{tab:flat} reports the complete family-level totals
rather than only favorable families.

\begin{table}[t]
\centering
\small
\begin{tabular}{lrrr}
\toprule
\textbf{Family} & \textbf{Evaluations} & \textbf{Invalid FLAT} & \textbf{False roots} \\
\midrule
F1 independent & 36 & 0 & 0 \\
F2 conjunction & 45 & 45 & 651 \\
F3 alternatives & 33 & 0 & 0 \\
F4 shared support & 33 & 33 & 2,022 \\
F5 irrelevant evidence & 36 & 0 & 0 \\
F6 mixed composition & 27 & 27 & 636 \\
F7 infeasible & 30 & 30 & 456 \\
F8 requirement expansion & 27 & 0 & 0 \\
\midrule
Total & 267 & 135 & 3,765 \\
\bottomrule
\end{tabular}
\caption{FLAT false coverage under the prespecified pairwise AND-erasure
projection. Invalid means the selected support failed closure under the
original graph.}
\label{tab:flat}
\end{table}

The prespecified pairwise projection is unsound relative to the original
derivation semantics when it converts required conjunctions into alternatives.
This result is specific to that projection, not a general comparison between
graphs and metadata.

\subsection{Support reduction}

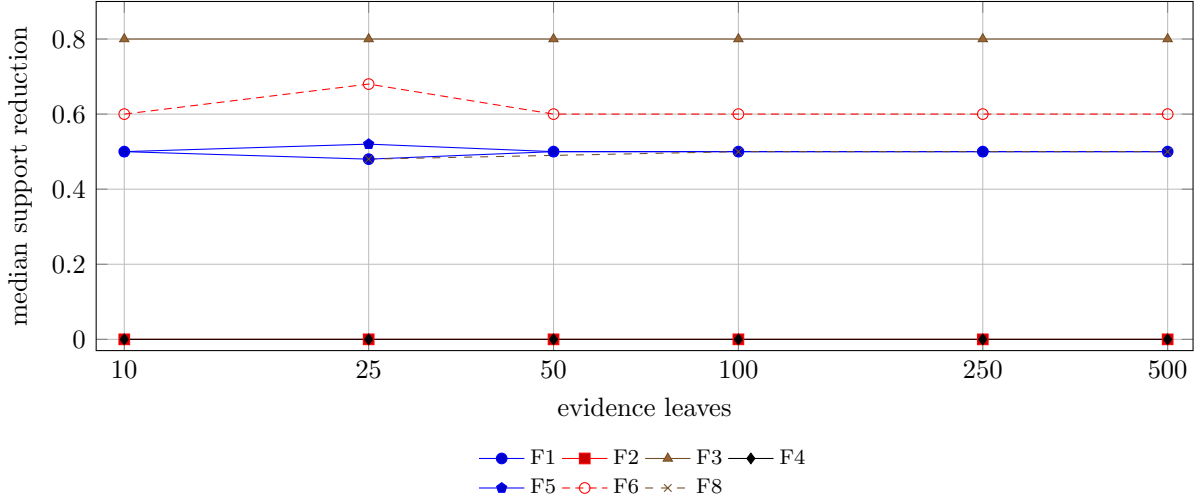
\begin{figure}[t]
\centering
\begin{tikzpicture}
\begin{axis}[
  width=0.94\linewidth,height=6.2cm,
  xmode=log,log basis x=10,
  xmin=9,xmax=550,ymin=-0.03,ymax=0.9,
  xlabel={evidence leaves},ylabel={median support reduction},
  xtick={10,25,50,100,250,500},xticklabels={10,25,50,100,250,500},
  grid=major,legend columns=4,
  legend style={font=\scriptsize,at={(0.5,-0.25)},anchor=north,draw=none},
  tick label style={font=\small},label style={font=\small}
]
\addplot+[mark=*] coordinates {(10,.50)(25,.48)(50,.50)(100,.50)(250,.50)(500,.50)}; \addlegendentry{F1}
\addplot+[mark=square*] coordinates {(10,0)(25,0)(50,0)(100,0)(250,0)(500,0)}; \addlegendentry{F2}
\addplot+[mark=triangle*] coordinates {(10,.80)(25,.80)(50,.80)(100,.80)(250,.80)(500,.80)}; \addlegendentry{F3}
\addplot+[mark=diamond*] coordinates {(10,0)(25,0)(50,0)(100,0)(250,0)(500,0)}; \addlegendentry{F4}
\addplot+[mark=pentagon*] coordinates {(10,.50)(25,.52)(50,.50)(100,.50)(250,.50)(500,.50)}; \addlegendentry{F5}
\addplot+[mark=o] coordinates {(10,.60)(25,.68)(50,.60)(100,.60)(250,.60)(500,.60)}; \addlegendentry{F6}
\addplot+[mark=x,dashed] coordinates {(25,.48)(100,.50)(500,.50)}; \addlegendentry{F8}
\end{axis}
\end{tikzpicture}
\caption{Median unit-cost support reduction versus retaining all evidence. F7
is omitted because no sufficient support exists for its roots. F8 has only three prespecified size
points.}
\label{fig:reduction}
\end{figure}

Figure~\ref{fig:reduction} shows median support reduction against \allmethod{}
at the frozen primary size points. The metric is
$1-|S|/|E|$ and is reported only for valid proven compositional optima. F1,
F3, F5, F6, and F8 have family-level medians of 0.50, 0.80, 0.50, 0.60, and
0.50. F2 and F4 have median zero because their frozen skeletons require all
leaves. Thus the benchmark includes both substantial and null reductions.

These reductions are properties of generated support cardinality. They do not
measure lower token use, verification runtime, money, or human effort.

\subsection{Exact agreement and exhaustive-search limit}

The mandatory n=10 gate passed for 21 of 21 instances. Across the optional
exact checks, 111 of 126 method-instance oracles completed, and every completed
check agreed with CP-SAT. Fifteen n=20 compositional checks timed out: all three
seeds in F2, F3, F4, F6, and F7. The exact flat oracle completed every n=20
instance. No CP-SAT-selected support failed the independent closure check.

\begin{table}[t]
\centering
\small
\begin{tabular}{crrr}
\toprule
\textbf{Size} & \textbf{Graph instances} & \textbf{Both methods complete} & \textbf{Compositional timeouts} \\
\midrule
10 & 21 & 21 & 0 \\
15 & 21 & 21 & 0 \\
20 & 21 & 6 & 15 \\
\midrule
Method-instance total & 63 & \multicolumn{2}{c}{111/126 completed} \\
\bottomrule
\end{tabular}
\caption{Exact-oracle completion. At n=20, compositional enumeration completed
only for F1 and F5; every completed exact result agreed with CP-SAT.}
\label{tab:oracle}
\end{table}

This crossover motivates a standard combinatorial optimizer for larger
prespecified graphs. It is not a new complexity result: minimum support already contains
known NP-hard special cases.

\subsection{Runtime behavior}

CP-SAT proved 234 of 267 compositional task evaluations optimal and 30
infeasible, all from F7. For the remaining three evaluations, it returned
closure-valid supports but did not prove optimality within 60 seconds.

\begin{figure}[t]
\centering
\begin{tikzpicture}
\begin{semilogyaxis}[
  width=0.94\linewidth,height=6.4cm,
  xmin=8,xmax=520,ymin=0.08,ymax=30,
  xlabel={evidence leaves},ylabel={median CP-SAT solve time (ms)},
  xtick={10,25,50,100,250,500},
  grid=major,legend columns=4,
  legend style={font=\scriptsize,at={(0.5,-0.25)},anchor=north,draw=none},
  tick label style={font=\small},label style={font=\small}
]
\addplot+[mark=*] coordinates {(10,.2125)(25,.243458)(50,.402791)(100,.695084)(250,1.488125)(500,2.883084)}; \addlegendentry{F1}
\addplot+[mark=square*] coordinates {(10,.2725)(25,.431583)(50,.743792)(100,1.387625)(250,3.278)(500,6.22125)}; \addlegendentry{F2}
\addplot+[mark=triangle*] coordinates {(10,.727541)(25,1.312)(50,2.176083)(100,3.865062)(250,9.680083)(500,19.385084)}; \addlegendentry{F3}
\addplot+[mark=diamond*] coordinates {(10,.251584)(25,.438875)(50,.720875)(100,1.369)(250,3.420584)(500,7.272875)}; \addlegendentry{F4}
\addplot+[mark=pentagon*] coordinates {(10,.202333)(25,.2455)(50,.375625)(100,.671709)(250,1.524667)(500,2.863292)}; \addlegendentry{F5}
\addplot+[mark=o] coordinates {(10,.293417)(25,.46275)(50,.87075)(100,1.668354)(250,3.9675)(500,8.441167)}; \addlegendentry{F6}
\addplot+[mark=x,dashed] coordinates {(10,.105583)(25,.20825)(50,.423875)(100,.786792)(250,1.919625)(500,3.881458)}; \addlegendentry{F7}
\end{semilogyaxis}
\end{tikzpicture}
\caption{Median CP-SAT solve time for the prespecified primary size sweep. F7
times measure proofs of infeasibility. The prespecified n=100 aggregation
includes the secondary cost points for F3 and F6. Timing is machine- and
encoding-specific.}
\label{fig:time}
\end{figure}
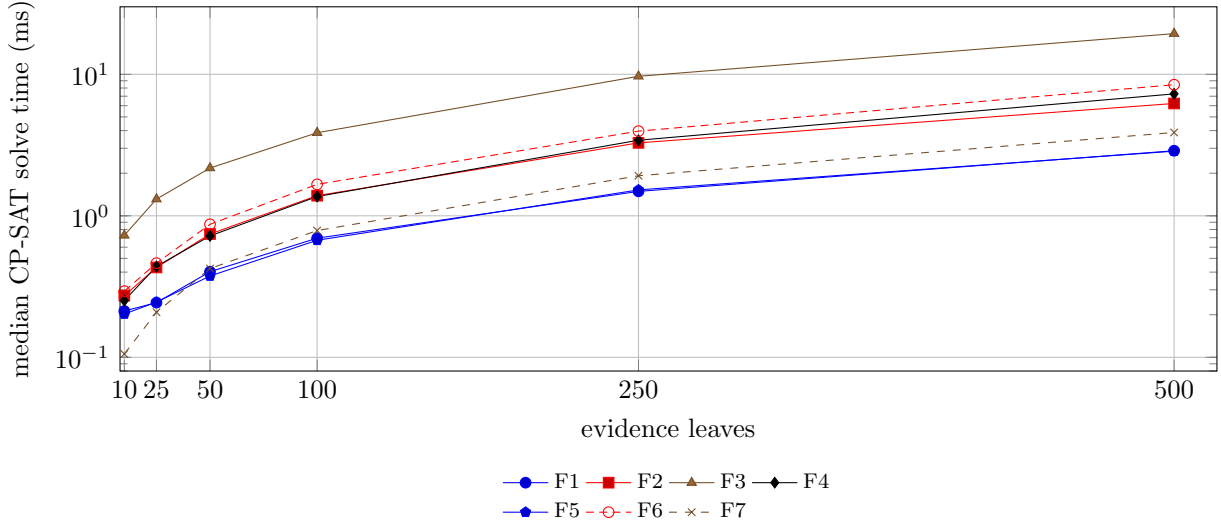

At the prespecified n=500 size points, median compositional solve time was below
20 ms for every family on the Apple M1 Pro test machine. Figure~\ref{fig:time}
shows the prespecified family/size medians. Raw evidence count is not sufficient
to predict difficulty: in the separate F3 structural sweep, all three n=100
instances with ten alternatives per target reached 60 seconds. Each returned a
valid cost-10 support, but none proved optimality and the best bound remained
zero. The prespecified F3 n=500 instances, which have five alternatives per target,
were proven optimal.

The hard F3 condition shows that alternative derivation structure can dominate
raw size under this generator and encoding. The data do not support a universal
scaling law or industrial-scale claim.

\section{Implementation}

The reported experiments evaluate envelope-selection semantics and optimization over
frozen graphs independently of the coding-agent runs that produced the underlying
software and evidence. The bounded research
implementation separates representation, validation,
semantics, and optimization. Immutable Python dataclasses represent evidence,
obligations, rules, and task requirements. Evidence carries a semantic type,
context identifier, assumptions, and integer cost. The existing assurance
record can serialize leaves and bind them to artifact, checker, scope, and
provenance; it is supporting infrastructure rather than the central result.

NetworkX represents the inference graph and checks node and edge
types, references, contexts, rule arity, costs, and acyclicity. The closure
validator performs deterministic forward chaining and records derivation
provenance. An exhaustive optimizer returns feasibility, all equal-cost optima,
and inclusion-minimal supports for bounded instances. The independent closure
validator checks every support returned by the CP-SAT adapter.

The scale generator uses a SHA-256 counter construction rather than platform
pseudorandomness. Canonical manifests bind family, parameters, seed, costs,
nodes, edges, roots, and graph hashes. All 249 manifests regenerated byte for
byte, and a second selected-support run reproduced the timing-free results.
The pinned environment used Python 3.12.14, NetworkX 3.6.1, and OR-Tools
9.15.6755. These implementation facts support reproducibility, not semantic
soundness of hand-specified obligations or rules.

\section{Discussion}

\subsection{Long-horizon use}

A long-horizon coding agent could request an envelope before modifying a component,
restoring only the evidence and derivations needed for the supplied task rather than
loading all prior engineering context or relying on current CI status. The
envelope would expose the evidence leaves and derivations required by supplied
task roots. A missing root would produce an explicit report that no sufficient envelope exists, instead
of unsupported reassurance. This use may reduce irrelevant context or repeated
checking, but those benefits have not been measured.

\subsection{Residual obligations and evolution}

The semantic/verification context $M$ changes what remains to establish. A
verified static judgment may discharge one root while leaving trace or
environment obligations external. Requirement evolution changes the roots even
when the artifact and prior evidence remain fixed, as the Pong graph shows.
Artifact evolution introduces a separate applicability problem. Changes to
source, dependencies, compiler, runtime, or assumptions may leave evidence
applicable, require its transport, make applicability uncertain, or invalidate
it.

Regression selection, change-impact analysis, incremental verification, proof
reuse, and verification caching already address parts of that lifecycle
\cite{rothermel,regverify,cacheverification,proofrepair}. Future work should
integrate their dependency analyses with typed assurance support rather than
claim a new invalidation algorithm.

\subsection{The unsolved input problem}

Envelope validity depends on a suitable $R(P,t,M)$ and sound rules. Candidate
roots may come from specifications, requirements links, program slicing,
dependency analysis, user clarification, or agent reasoning. Intent
formalization and executable-specification work directly address this boundary
\cite{intent,ticoder,harel}. This study supplies roots manually; future
evaluations should test automatic root construction and agent decisions made
with the selected support.

\subsection{Broader impact}

As a defensive mechanism, a task-conditioned envelope is intended to reduce unsupported
acceptance of automated software change: a required obligation that current evidence
cannot derive surfaces as an explicit infeasible result rather than as unsupported
reassurance. Its principal risk is misplaced trust, since an envelope certifies only
that the selected evidence derives the supplied obligations under the supplied rules,
not that those obligations are complete or that the underlying judgments are sound;
treating a sufficient envelope as a correctness guarantee would overstate what it
establishes. The formulation adds no generative capability and operates over evidence
already produced by prior engineering work, so we identify no further dual-use concern.

\section{Threats to Validity}

\paragraph{Construct validity.}
The obligation sets and inference rules are hand-specified. They may omit real
requirements or encode a bridge too strongly. \flatmethod{} is one prespecified
pairwise AND-erasure abstraction, not every non-compositional evidence system.
Unit and small integer costs measure support selection, not real acquisition,
context, or risk.

\paragraph{Internal validity.}
The deterministic generator uses hand-designed graph families that isolate
specific structures. This improves control but can favor the phenomena under
study. F7 infeasibility is constructed. Mandatory exhaustive checks,
independent closure validation, pinned dependencies, canonical manifests, and
deterministic reruns reduce implementation risk but do not establish the
semantic truth of supplied rules.

\paragraph{External validity.}
Rust, IronBlocks, and Pong yield small provenance-derived graphs, not industrial
assurance networks. The 249 scale instances are synthetic and stop at 500
evidence leaves. The evaluation includes software and evidence produced by prior
coding-agent runs, but does not test whether a downstream coding agent performs better
when given the selected envelope. Pong's numerical
evidence is documented and reconstructed rather than freshly reproduced.

\paragraph{Conclusion validity.}
Timing comes from one run per prespecified graph and method on an Apple M1 Pro,
so it is machine-specific. Three feasible CP-SAT runs did not prove optimality,
and exact enumeration timed out on selected n=20 instances. Optimum and
reduction claims exclude those supports.

The benchmark establishes computational and representational behavior when
task obligations and inference rules are supplied. It does not establish that
real coding agents infer those obligations correctly or perform better when
given the resulting envelope.

\section{Related Work}

\subsection{Assurance structures and evolution}

The Goal Structuring Notation (GSN) and assurance cases structure claims,
arguments, context, and evidence \cite{gsn,kellypatterns}. The Structured Assurance Case
Metamodel (SACM) standardizes a machine-readable metamodel for assurance arguments and
artifacts \cite{sacm}; OpenCert and AMASS provide
lifecycle repositories and tooling \cite{opencert,amass,amassarch}. ACCESS and
Isabelle/SACM connect evolving cases to engineering artifacts and formal proof
\cite{access,isabellesacm}. Dynamic safety cases, perpetual assurance, and dynamic-assurance lifecycles treat
evidence and argument state as evolving \cite{dynamicsafety,perpetual,dynamicassurance}. Our
graph can be represented as a narrow assurance/proof slice. Its specific focus
is cost-aware support selected for supplied roots of a future coding task, not
a competing assurance metamodel.

\subsection{Selection, reuse, and minimum support}

Regression-test selection and industrial test-impact analysis select tests
affected by change \cite{rothermel,googletestselection}. Regression
verification, fine-grained caching, proof repositories, and proof repair reuse
or update formal work across versions
\cite{regverify,cacheverification,proofrepo,proofrepair}. Program slicing
selects code relevant to a semantic criterion \cite{weiser}. These techniques
can help construct task roots or determine evidence applicability. The current
study treats task roots and evidence applicability as inputs and solves only the
support-selection step.

Weighted set cover is the independent-coverage special case, with classical
approximation bounds \cite{feige}. Directed hypergraphs and AND/OR proof graphs
represent conjunctive premises and alternatives \cite{gallo}; logic-based
abduction studies minimum explanations under a theory \cite{eiter}. The
optimization is therefore known. Our implementation applies it to typed,
artifact- and context-indexed software evidence and independently validates
selected supports.

\subsection{Program logics and enforcement}

Assume--guarantee reasoning and contract theories compose component obligations
\cite{contracts}. Separation logic, refinement types, effect systems,
behavioral types, temporal logic, model checking, proof-carrying code, and
certified compilation establish other typed judgments
\cite{reynolds,liquid,effects,sessiontypes,pnueli,modelchecking,pcc,compcert}.
The envelope records their typed conclusions and justified bridges; it does not
unify their semantics or introduce a verification method.

Runtime monitoring and enforcement distinguish observation from intervention
and characterize enforceable policies \cite{schneider,editautomata,falcone}.
IronBlocks instantiates this known distinction as a supplied conjunction between
detection and commitment mediation.

\subsection{Provenance, intent, and coding agents}

in-toto and Supply-chain Levels for Software Artifacts (SLSA) capture
software-supply-chain provenance, while Sigstore binds software signatures to identities
and transparency records \cite{intoto,slsa,sigstore}. They establish that artifact-bound
provenance, attestation, and signing are not new. The envelope adds task-selected behavioral obligations
and derivation structure; it does not replace supply-chain provenance.

Intent formalization and test-driven or executable specifications study the
gap between informal requests and checkable requirements
\cite{intent,ticoder,harel}. ReAct, KISS Sorcar, and AlphaVerus illustrate
tool-using or verifier-guided agent workflows \cite{react,kiss,alphaverus}, while
DafnyCOMP studies compositional formal verification in current models \cite{dafnycomp}. Our evaluation starts with supplied
obligations and asks which existing evidence supports them; evaluating an agent
consumer remains future work.

\paragraph{Agent context selection.} Long-horizon language-model agents must decide
which prior information to expose within a bounded working context. MemGPT manages
information across memory tiers for persistent agents, Repoformer selectively retrieves
repository context when it is expected to help code generation, and SWE-agent shows that
the observations and interfaces exposed to a coding agent affect its behavior
\cite{packer2023memgpt,wu2024repoformer,yang2024sweagent}. An assurance envelope selects
a different object: machine-produced engineering evidence whose explicit derivations
cover supplied software obligations. We do not yet measure whether providing that
selected evidence improves downstream coding-agent performance.

\section{Conclusion}

For a supplied software-change obligation set, the relevant assurance context
can be represented as a support problem over typed evidence and explicit
composition rules. A task-conditioned assurance envelope is sufficient when
its closure derives every required root, and minimum-cost selection avoids
retaining unnecessary leaves under the chosen cost model.

Graphs constructed from preserved software evidence exhibit task-dependent
support, infeasibility, conjunctive composition, and requirement expansion. In
the separate 249-instance synthetic benchmark, the pairwise baseline loses
conjunction structure, support reduction ranges from null to substantial, and
every completed exact check agrees with CP-SAT under bounded conditions. The
open problems are to construct and maintain $R(P,t,M)$ as code, context, and
intent evolve, and then to evaluate whether coding agents benefit from consuming the
resulting task-conditioned assurance envelope.

{\small
\bibliographystyle{plain}
\bibliography{assurance-envelopes}

\begin{thebibliography}{10}

\bibitem{alphaverus}
Pranjal Aggarwal, Bryan Parno, and Sean Welleck.
\newblock {AlphaVerus}: Bootstrapping formally verified code generation through
  self-improving translation and treefinement.
\newblock In {\em Proceedings of the International Conference on Machine
  Learning}, volume 267 of {\em Proceedings of Machine Learning Research},
  pages 587--615, 2025.

\bibitem{batty}
Mark Batty, Scott Owens, Susmit Sarkar, Peter Sewell, and Tjark Weber.
\newblock Mathematizing {C++} concurrency.
\newblock In {\em Proceedings of POPL}, pages 55--66, 2011.

\bibitem{proofrepo}
Richard Bubel, Ferruccio Damiani, Reiner H{\"a}hnle, Einar~Broch Johnsen, Olaf
  Owe, Ina Schaefer, and Ingrid~Chieh Yu.
\newblock Proof repositories for compositional verification of evolving
  software systems---managing change when proving software correct.
\newblock {\em Transactions on Foundations for Mastering Change}, 1:130--156,
  2016.

\bibitem{dynamicassurance}
Radu Calinescu, Danny Weyns, Simos Gerasimou, M.~Usman Iftikhar, Ibrahim Habli,
  and Tim Kelly.
\newblock Engineering trustworthy self-adaptive software with dynamic assurance
  cases.
\newblock {\em IEEE Transactions on Software Engineering}, 44(11):1039--1069,
  2018.

\bibitem{contracts}
Chris Chilton, Bengt Jonsson, and Marta Kwiatkowska.
\newblock Compositional assume--guarantee reasoning for input/output component
  theories.
\newblock {\em Science of Computer Programming}, 91:115--137, 2014.

\bibitem{modelchecking}
Edmund~M. Clarke, Orna Grumberg, and Doron~A. Peled.
\newblock {\em Model Checking}.
\newblock MIT Press, 1999.

\bibitem{amass}
Jos{\'e}~Luis de~la Vara, Alejandra Ruiz, and Ga{\"e}l Blondelle.
\newblock Assurance and certification of cyber--physical systems: The {AMASS}
  open source ecosystem.
\newblock {\em Journal of Systems and Software}, 171:110812, 2021.

\bibitem{dynamicsafety}
Ewen Denney, Ganesh~J. Pai, and Ibrahim Habli.
\newblock Dynamic safety cases for through-life safety assurance.
\newblock In {\em Proceedings of the 37th International Conference on Software
  Engineering}, volume~2, pages 587--590, 2015.

\bibitem{opencert}
{Eclipse Foundation}.
\newblock Eclipse opencert.
\newblock Open-source project, release 1.0.0,
  \url{https://projects.eclipse.org/projects/polarsys.opencert}, 2019.

\bibitem{eiter}
Thomas Eiter and Georg Gottlob.
\newblock The complexity of logic-based abduction.
\newblock {\em Journal of the ACM}, 42(1):3--42, 1995.

\bibitem{falcone}
Yli{\`e}s Falcone, Laurent Mounier, Jean-Claude Fernandez, and Jean-Luc
  Richier.
\newblock Runtime enforcement monitors: Composition, synthesis, and enforcement
  abilities.
\newblock {\em Formal Methods in System Design}, 38(3):223--262, 2011.

\bibitem{feige}
Uriel Feige.
\newblock A threshold of $\ln n$ for approximating set cover.
\newblock {\em Journal of the ACM}, 45(4):634--652, 1998.

\bibitem{isabellesacm}
Simon Foster, Yakoub Nemouchi, Mario Gleirscher, Ran Wei, and Tim Kelly.
\newblock Integration of formal proof into unified assurance cases with
  {Isabelle/SACM}.
\newblock {\em Formal Aspects of Computing}, 33:855--884, 2021.

\bibitem{gallo}
Giorgio Gallo, Giustino Longo, Stefano Pallottino, and Sang Nguyen.
\newblock Directed hypergraphs and applications.
\newblock {\em Discrete Applied Mathematics}, 42(2--3):177--201, 1993.

\bibitem{regverify}
Benny Godlin and Ofer Strichman.
\newblock Regression verification: Proving the equivalence of similar programs.
\newblock {\em Software Testing, Verification and Reliability}, 23(3):241--258,
  2013.

\bibitem{linearizability}
Maurice~P. Herlihy and Jeannette~M. Wing.
\newblock Linearizability: A correctness condition for concurrent objects.
\newblock {\em ACM Transactions on Programming Languages and Systems},
  12(3):463--492, 1990.

\bibitem{sessiontypes}
Kohei Honda, Vasco~T. Vasconcelos, and Makoto Kubo.
\newblock Language primitives and type discipline for structured
  communication-based programming.
\newblock In {\em Programming Languages and Systems}, pages 122--138, 1998.

\bibitem{rustbelt}
Ralf Jung, Jacques-Henri Jourdan, Robbert Krebbers, and Derek Dreyer.
\newblock {RustBelt}: Securing the foundations of the rust programming
  language.
\newblock {\em Proceedings of the ACM on Programming Languages},
  2(POPL):66:1--66:34, 2018.

\bibitem{kellypatterns}
Tim Kelly and John McDermid.
\newblock Safety case patterns---reusing successful arguments.
\newblock In {\em IEE Colloquium on Understanding Patterns and Their
  Application to System Engineering}, 1998.

\bibitem{intent}
Shuvendu~K. Lahiri.
\newblock Intent formalization: A grand challenge for reliable coding in the
  age of {AI} agents.
\newblock {\em arXiv preprint arXiv:2603.17150}, 2026.

\bibitem{ticoder}
Shuvendu~K. Lahiri, Sarah Fakhoury, Aaditya Naik, Georgios Sakkas, Saikat
  Chakraborty, Madanlal Musuvathi, Piali Choudhury, Curtis von Veh,
  Jeevana~Priya Inala, Chenglong Wang, and Jianfeng Gao.
\newblock Interactive code generation via test-driven user-intent
  formalization.
\newblock {\em arXiv preprint arXiv:2208.05950}, 2022.

\bibitem{cacheverification}
K.~Rustan~M. Leino and Valentin W{\"u}stholz.
\newblock Fine-grained caching of verification results.
\newblock In {\em Computer Aided Verification}, pages 380--397, 2015.

\bibitem{googletestselection}
Claire Leong, Abhayendra Singh, John Micco, Mike Papadakis, and Yves Le~Traon.
\newblock Assessing transition-based test selection algorithms at google.
\newblock In {\em Proceedings of the 41st International Conference on Software
  Engineering: Software Engineering in Practice}, 2019.

\bibitem{compcert}
Xavier Leroy.
\newblock Formal verification of a realistic compiler.
\newblock {\em Communications of the ACM}, 52(7):107--115, 2009.

\bibitem{editautomata}
Jay Ligatti, Lujo Bauer, and David Walker.
\newblock Edit automata: Enforcement mechanisms for run-time security policies.
\newblock {\em International Journal of Information Security}, 4(1--2):2--16,
  2005.

\bibitem{effects}
John~M. Lucassen and David~K. Gifford.
\newblock Polymorphic effect systems.
\newblock In {\em Proceedings of POPL}, pages 47--57, 1988.

\bibitem{pcc}
George~C. Necula.
\newblock Proof-carrying code.
\newblock In {\em Proceedings of POPL}, pages 106--119, 1997.

\bibitem{sigstore}
Zachary Newman, John~Speed Meyers, and Santiago Torres-Arias.
\newblock Sigstore: Software signing for everybody.
\newblock In {\em Proceedings of the 2022 ACM SIGSAC Conference on Computer and
  Communications Security}, pages 2353--2367, 2022.

\bibitem{sacm}
{Object Management Group}.
\newblock Structured assurance case metamodel ({SACM}), version 2.3.
\newblock OMG Formal Specification, \url{https://www.omg.org/spec/SACM/2.3},
  2023.

\bibitem{slsa}
{OpenSSF SLSA Working Group}.
\newblock Supply-chain levels for software artifacts ({SLSA}), version 1.2.
\newblock Official specification, \url{https://slsa.dev/spec/v1.2/}, 2025.

\bibitem{packer2023memgpt}
Charles Packer, Sarah Wooders, Kevin Lin, Vivian Fang, Shishir~G. Patil, Ion
  Stoica, and Joseph~E. Gonzalez.
\newblock {MemGPT}: Towards llms as operating systems.
\newblock {\em arXiv preprint arXiv:2310.08560}, 2023.

\bibitem{pnueli}
Amir Pnueli.
\newblock The temporal logic of programs.
\newblock In {\em 18th IEEE Symposium on Foundations of Computer Science},
  pages 46--57, 1977.

\bibitem{harel}
Dezhi Ran, Mengzhou Wu, Yuan Cao, Assaf Marron, David Harel, and Tao Xie.
\newblock An infrastructure software perspective toward computation offloading
  between executable specifications and foundation models.
\newblock {\em Science China Information Sciences}, 68(4):146101, 2025.

\bibitem{reynolds}
John~C. Reynolds.
\newblock Separation logic: A logic for shared mutable data structures.
\newblock In {\em Proceedings of LICS}, pages 55--74, 2002.

\bibitem{proofrepair}
Talia Ringer, RanDair Porter, Nathaniel Yazdani, John Leo, and Dan Grossman.
\newblock Proof repair across type equivalences.
\newblock In {\em Proceedings of PLDI}, pages 112--127, 2021.

\bibitem{liquid}
Patrick~M. Rondon, Ming Kawaguchi, and Ranjit Jhala.
\newblock Liquid types.
\newblock In {\em Proceedings of PLDI}, pages 159--169, 2008.

\bibitem{rothermel}
Gregg Rothermel and Mary~Jean Harrold.
\newblock A safe, efficient regression test selection technique.
\newblock {\em ACM Transactions on Software Engineering and Methodology},
  6(2):173--210, 1997.

\bibitem{amassarch}
Alejandra Ruiz, Barbara Gallina, Jos{\'e}~Luis de~la Vara, Silvia Mazzini, and
  Hu{\'a}scar Espinoza.
\newblock Architecture-driven, multi-concern and seamless assurance and
  certification of cyber-physical systems.
\newblock In {\em SAFECOMP Workshops}, pages 311--321, 2016.

\bibitem{schneider}
Fred~B. Schneider.
\newblock Enforceable security policies.
\newblock {\em ACM Transactions on Information and System Security},
  3(1):30--50, 2000.

\bibitem{gsn}
{SCSC Assurance Case Working Group}.
\newblock Goal structuring notation community standard, version 3.
\newblock SCSC-141C, 2021.

\bibitem{kiss}
Koushik Sen.
\newblock {KISS Sorcar}: A stupidly-simple general-purpose and software
  engineering {AI} assistant.
\newblock {\em arXiv preprint arXiv:2604.23822}, 2026.

\bibitem{intoto}
Santiago Torres-Arias, Hammad Afzali, Trishank~Karthik Kuppusamy, Reza
  Curtmola, and Justin Cappos.
\newblock in-toto: Providing farm-to-table guarantees for bits and bytes.
\newblock In {\em 28th USENIX Security Symposium}, pages 1393--1410, 2019.

\bibitem{access}
Ran Wei, Simon Foster, Haitao Mei, Fang Yan, Ruizhe Yang, Ibrahim Habli, Colin
  O'Halloran, Nick Tudor, Tim Kelly, and Yakoub Nemouchi.
\newblock {ACCESS}: Assurance case centric engineering of safety-critical
  systems.
\newblock {\em Journal of Systems and Software}, 213:112034, 2024.

\bibitem{weiser}
Mark Weiser.
\newblock Program slicing.
\newblock In {\em Proceedings of the 5th International Conference on Software
  Engineering}, pages 439--449, 1981.

\bibitem{perpetual}
Danny Weyns, Nelly Bencomo, Radu Calinescu, Javier C{\'a}mara, Carlo Ghezzi,
  Vincenzo Grassi, Lars Grunske, Paola Inverardi, Jean-Marc J{\'e}z{\'e}quel,
  Sam Malek, Raffaela Mirandola, Marco Mori, and Giordano Tamburrelli.
\newblock Perpetual assurances for self-adaptive systems.
\newblock {\em arXiv preprint arXiv:1903.04771}, 2019.

\bibitem{wu2024repoformer}
Di~Wu, Wasi~Uddin Ahmad, Dejiao Zhang, Murali~Krishna Ramanathan, and Xiaofei
  Ma.
\newblock Repoformer: Selective retrieval for repository-level code completion.
\newblock In {\em Proceedings of the 41st International Conference on Machine
  Learning}, volume 235 of {\em Proceedings of Machine Learning Research},
  pages 53270--53290. PMLR, 2024.

\bibitem{dafnycomp}
Xu~Xu, Xin Li, Xingwei Qu, Jie Fu, and Binhang Yuan.
\newblock Local success does not compose: Benchmarking large language models
  for compositional formal verification.
\newblock In {\em International Conference on Learning Representations}, 2026.

\bibitem{yang2024sweagent}
John Yang, Carlos~E. Jimenez, Alexander Wettig, Kilian Lieret, Shunyu Yao,
  Karthik Narasimhan, and Ofir Press.
\newblock {SWE-agent}: Agent-computer interfaces enable automated software
  engineering.
\newblock In {\em Advances in Neural Information Processing Systems},
  volume~37, 2024.

\bibitem{react}
Shunyu Yao, Jeffrey Zhao, Dian Yu, Nan Du, Izhak Shafran, Karthik~R.
  Narasimhan, and Yuan Cao.
\newblock {ReAct}: Synergizing reasoning and acting in language models.
\newblock In {\em International Conference on Learning Representations}, 2023.

\end{thebibliography}
}

\end{document}